\documentclass[onecolumn,aps,prc,floatfix,superscriptaddress]{revtex4-1}
\usepackage{amsmath,bm}
\usepackage{graphicx}
\usepackage{epstopdf}
\usepackage{subfigure}
\usepackage{epsfig}
\usepackage{amsmath,amssymb,amsfonts}
\usepackage{color}
\usepackage[utf8]{inputenc}
\usepackage{verbatim}
\usepackage{array}
\usepackage{lineno}
\usepackage{hyperref}
\graphicspath{{figures/}}
\begin{document}

\title{Improved Pion–Kaon Identification in Heavy-Ion Collisions with a Two-Dimensional Transformation}

\author{Shaowei Lan}
\email{shaoweilan@pdsu.edu.cn}
\affiliation{School of Electrical and Mechanical Engineering, Pingdingshan University, 467000 Pingdingshan, China}

\author{Bijun Fan}
\email{bjfan@mails.ccnu.edu.cn}
\affiliation{Key Laboratory of Quark \& Lepton Physics (MOE) and Institute of Particle Physics, Central China Normal University, Wuhan 430079, China}

\author{Like Liu}
\email{likeliu@mails.ccnu.edu.cn}
\affiliation{Key Laboratory of Quark \& Lepton Physics (MOE) and Institute of Particle Physics, Central China Normal University, Wuhan 430079, China}


\date{\today}

\begin{abstract}

Accurate identification of charged pions and kaons is essential for precision measurements in relativistic heavy-ion collisions, but becomes increasingly challenging at intermediate and high transverse momentum because of the increasing overlap of the TPC and TOF particle identification responses. In this work, we quantitatively benchmark a two-dimensional shift-and-rotation PID procedure that combines the $n\sigma_{\pi}$ and mass-squared $(m^{2})$ information. The benchmark is performed using Au+Au collision events generated with the AMPT model, where the detector response is modeled using $p_{T}$-dependent PID parameters extracted from STAR data. Since the true particle identities and input yields are known in AMPT but are not used during the reconstruction, the procedure can be evaluated through a direct closure test that is not possible with experimental data alone.
The two-dimensional procedure is compared with a reference one-dimensional method consisting of an $|n\sigma_{\pi}|<$ 3 preselection followed by a Gaussian fit to the $m^{2}$ distribution. While both methods perform well when pion and kaon PID responses are well separated, the two-dimensional procedure provides improved yield closure as the PID distributions increasingly overlap at high $p_{T}$. The reconstructed-to-input yield ratio remains within 2\% of unity up to $p_T =$ 3 GeV/$c$, whereas the reference one-dimensional procedure shows a larger deviation. 
Reaction-plane $v_{2}$ closure tests show that both the reference one-dimensional and two-dimensional PID procedures reproduce the AMPT input $v_{2}$ within the statistical uncertainties over the studied $p_{T}$ range.
These results demonstrate that the two-dimensional procedure substantially improves the high-$p_{T}$ yield reconstruction without introducing a detectable bias in the extracted $v_{2}$ under the STAR-like detector response considered here.

\end{abstract}

\maketitle

\section{Introduction}
\label{sec:introduction}
Relativistic heavy-ion collisions, such as those conducted at the Relativistic Heavy Ion Collider (RHIC) and the Large Hadron Collider (LHC), accelerate heavy nuclei to nearly the speed of light and collide them to recreate the extreme energy density and temperature conditions of the early universe. These collisions offer a unique opportunity to study the properties of a novel state of matter known as the quark-gluon plasma (QGP), in which these quarks and gluons are no longer confined within hadrons~\cite{Braun-Munzinger:2007edi,Shuryak:1978ij}.
The QGP medium is studied indirectly by analyzing the properties of the hadrons produced in the final state of the collisions. A variety of observables including collective flow, hadron yield distributions, and jet quenching patterns have been proposed and extensively studied to probe the transport and collective behavior of the QGP~\cite{Gyulassy:1990ye,Qin:2015srf,STAR:2015gge,STAR:2017sal,Chen:2024aom,Luo:2020pef,Bzdak:2019pkr}. Central to all these measurements is the accurate identification of particle species, as different hadrons carry distinct information about the system's evolution, freeze-out dynamics, and the underlying QCD processes. 

Among the hadrons commonly observed, charged pions ($\pi^\pm$), kaons ($K^\pm$), and protons ($p/ \bar{p}$) are of particular importance due to their relatively long lifetimes. These particles can be directly detected and identified on a track-by-track basis using detectors such as Time Projection Chamber (TPC)~\cite{Anderson:2003ur} and Time-of-Flight (TOF) systems~\cite{Geurts:2004fn}. The TPC provides measurements of the ionization energy loss ($dE/dx$), while the TOF system offers timing information that can be used to calculate the squared mass ($m^2$) of the particle. Together, these detectors enable particle identification (PID) in a wide range of transverse momentum ($p_T$)~\cite{STAR:2002eio}, although with increasing difficulty at high $p_T$ where the resolution of PID signals overlaps between species.
In contrast, hadrons originating from the decay of short-lived resonances or strange baryons, such as $\Lambda$, $\Xi$ , or $K_S^0$, have lifetimes too short to be directly measured. These particles are typically reconstructed through their weak decay topologies using secondary vertex identification techniques, which are sensitive to tracking resolution and background contamination. These particle reconstruction are performed using the KF Particle Finder package based on the Kalman Filter method~\cite{Kisel:2018nvd}. 
A persistent challenge in charged hadron identification is the separation of pions and kaons at intermediate and high $p_T$ (typically above 2.0 GeV/$c$). As the particle momentum increases, the expected TPC $dE/dx$ and TOF $m^{2}$ responses of pions and kaons become progressively less separated relative to the detector resolution~\cite{STAR:2008med}, resulting in increasing overlap of the PID distributions~\cite{Aguilar-Benitez:1991hzq}. Such overlap can reduce the efficiency of particle identification, increase cross-contamination between species, and introduce systematic uncertainties in measurements of identified particle spectra, anisotropic flow, and correlations. Improving pion-kaon separation in this kinematic region is therefore important for precision identified hadron measurements and is a key objective in ongoing and future heavy-ion collision experiments~\cite{Yang:2013yeb, CBM:2016kpk,Kekelidze:2016wkp}.


Two-dimensional PID strategies combining the TPC $n\sigma$ (normalized $dE/dx$)~\cite{Bichsel:2006cs} and TOF $m^{2}$ information have previously been employed in STAR identified-hadron analyses using real collision data~\cite{STAR:2013cow, STAR:2013ayu, STAR:2015rxv, STAR:2012och, STAR:2021yiu}. In particular, the transformation of the original $n\sigma-m^{2}$ plane into a shifted and rotated coordinate system has been used to improve pion-kaon separation when the individual PID responses strongly overlap~\cite{STAR:2013ayu,STAR:2021yiu}. The experimental applicability of such a procedure has therefore already been established in previous STAR analyses.
The purpose of the present work is complementary. Rather than introducing the two-dimensional procedure as a new experimental PID technique, we use a controlled Monte Carlo framework to quantify its performance relative to a reference one-dimensional PID procedure. The detector response is modeled using $p_{T}$-dependent PID parameters obtained from STAR data and applied to AMPT particles through Gaussian smearing. The AMPT particle identities are not used during the PID reconstruction and are retained only as a truth reference for the final closure comparison. This enables a direct quantitative assessment of the reconstructed yields and of possible PID-induced bias in elliptic flow measurements.

This paper is organized as follows. Section~\ref{sec:method} introduces the AMPT model, the STAR data calibrated detector response smearing, the reference one-dimensional PID procedure, and the two-dimensional shift and rotation method.
Section~\ref{sec:res} presents the quantitative yield closure comparison and the reaction-plane $v_{2}$ closure comparison.
Finally, a summary is given in Section~\ref{sec:sum}.

\section{Method}
\label{sec:method}

This simulation is performed using a multi-phase transport model (AMPT)~\cite{Lin:2004en} to generate Au+Au collision events at $\sqrt{s_{NN}}$ = 200 GeV. In this study, the string melting version of AMPT~\cite{Lin:2001zk} is employed, which includes four main components: fluctuating initial conditions from HIJING, partonic scatterings described by the Zhang’s Parton Cascade (ZPC), hadronization via a quark coalescence mechanism, and subsequent hadronic interactions modeled by the ART framework. The AMPT model has been extensively validated against experimental measurements at RHIC and LHC energies and has demonstrated its capability to reproduce a wide range of observables, including particle spectra and anisotropic flow coefficients~\cite{Lin:2014tya,Lan:2017nye, Liu:2025eml}.



\begin{figure}[htbp]
\centering
\centerline{\includegraphics[width=0.6\linewidth]{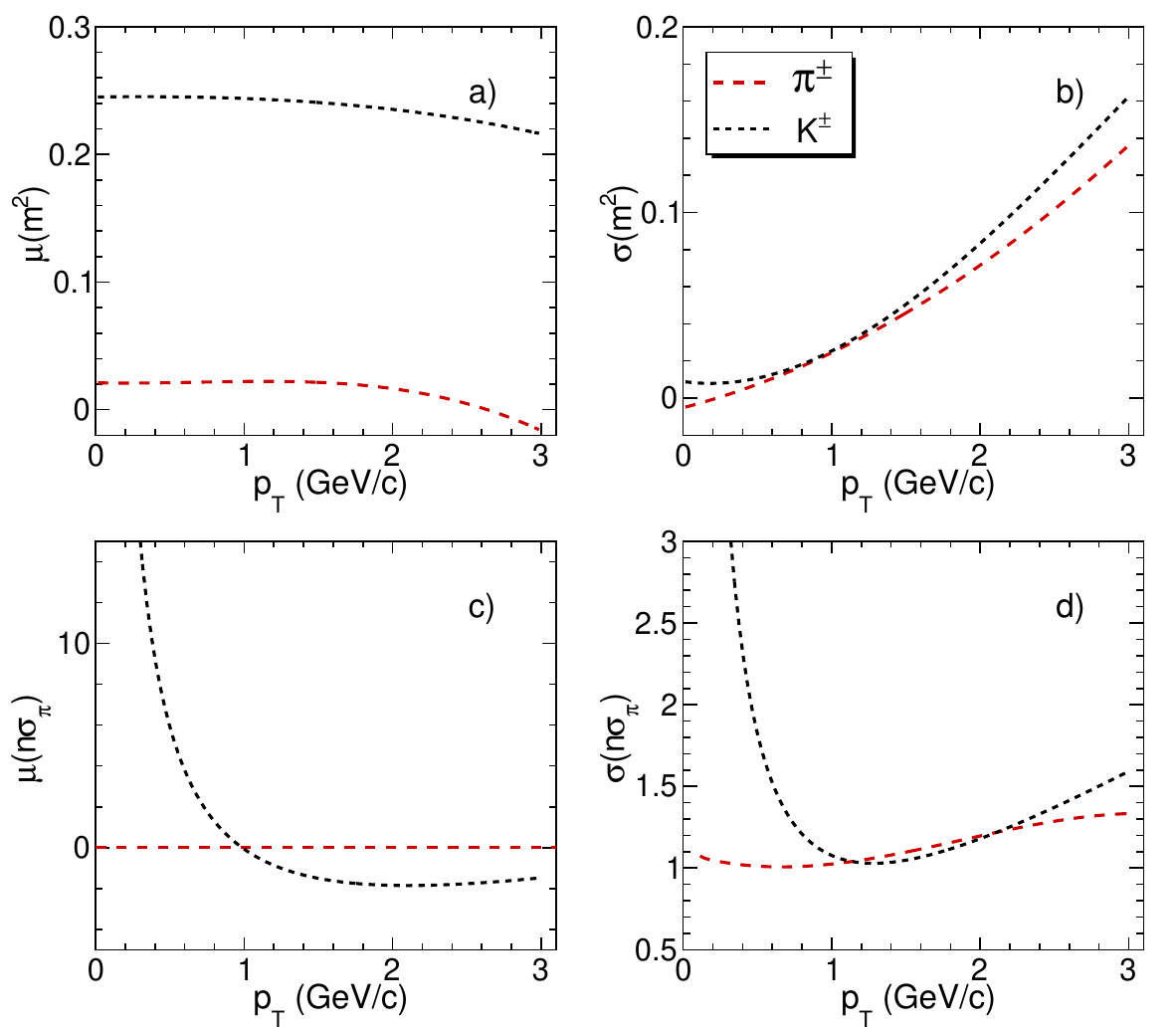}}
\caption {Transverse momentum ($p_T$) dependence of the pion and kaon PID response parameters used in the detector smearing. The parameters are obtained from STAR Au+Au data at $\sqrt{s_{NN}}$= 200 GeV and are used to generate the Gaussian $m^{2}$ and $n\sigma_{\pi}$ responses of the simulated pion and kaon samples. Panels (a) and (b) show the $p_{T}$-dependent mean ($\mu$) and width ($\sigma$) of the $m^{2}$ response, respectively, while panels (c) and (d) show the corresponding mean ($\mu$) and width ($\sigma$) of the $n\sigma_{\pi}$ response.}
\label{fig:fig1}
\end{figure}

The AMPT model provides the true particle species and momenta but does not include the finite PID resolution of an experimental detector.  To construct a controlled STAR-like PID response, the $m^{2}$ and $n\sigma_{\pi}$ distributions of charged pions and kaons are smeared using $p_{T}$-dependent mean values and widths extracted from STAR Au+Au data at $\sqrt{s_{NN}}$= 200 GeV. These parameters were obtained from the PID calibration procedure used in the corresponding STAR analyses~\cite{STAR:2015gge}, and the parameterizations adopted in the present study are shown in Fig.~\ref{fig:fig1}.
The variable $n\sigma_{\pi}$ is used as the common TPC PID coordinate for both species. Pions are centered near $n\sigma_{\pi}=0$, whereas kaons appear as a displaced distribution in the same variable; therefore, no independent $n\sigma_{K}$ coordinate is required for the present analysis. For each AMPT pion or kaon, the $m^{2}$ and $n\sigma_{\pi}$ observables are independently sampled from Gaussian distributions characterized by the corresponding $p_{T}$-dependent means and widths. The pion and kaon relative abundances are inherited directly from the AMPT events and are not artificially reweighted.
After the common kinematic selections and detector response are applied, the pion and kaon tracks are combined and analyzed without using the AMPT truth particle identity. The truth information is retained only for the final comparison between the reconstructed and input yields. 
The smeared sample provides a controlled environment in which the PID reconstruction is performed without truth labels, while the reconstruction bias can still be quantified directly using the known AMPT input.

\begin{figure}[htbp]
\centering
\centerline{\includegraphics[width=0.6\linewidth]{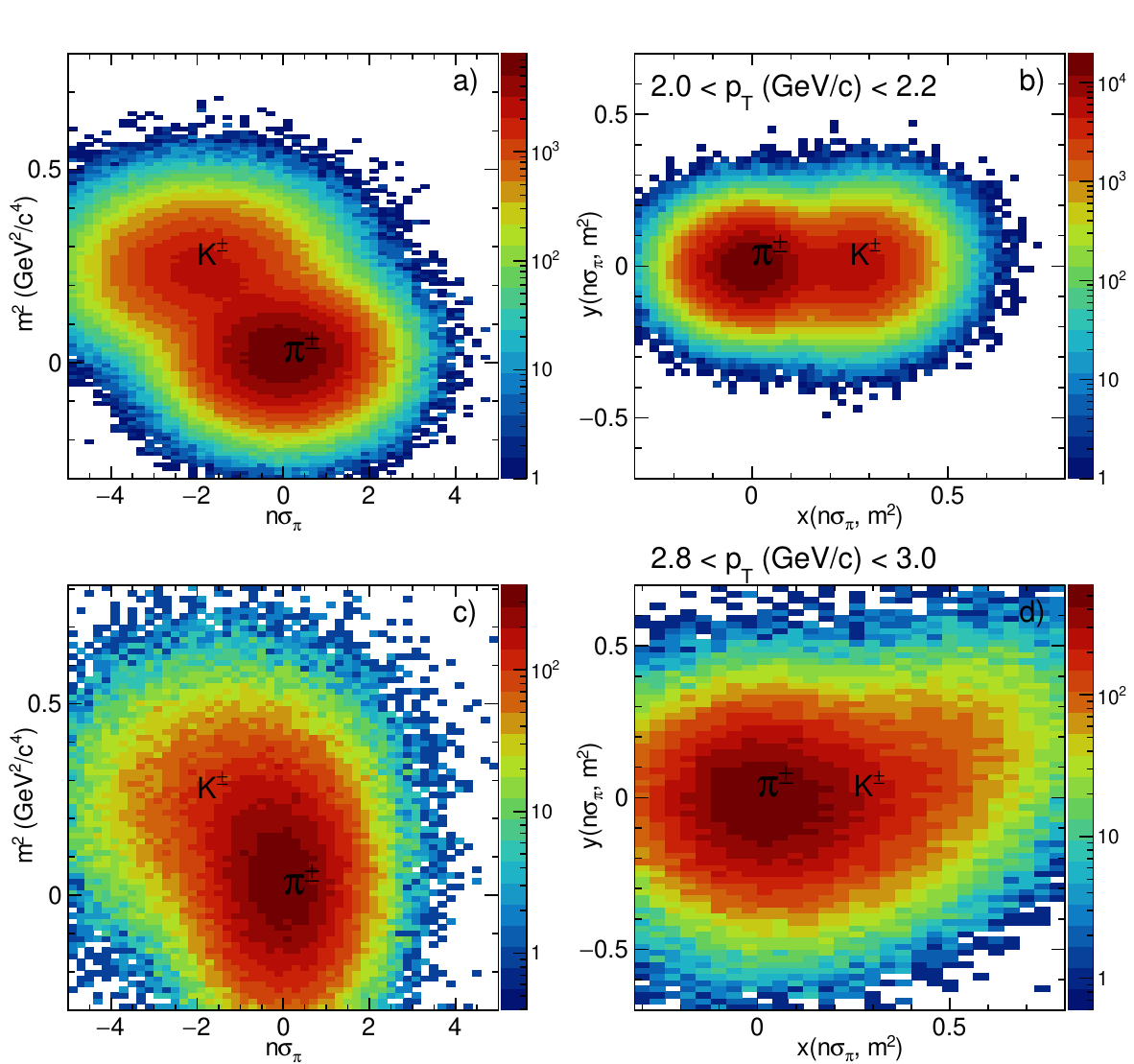}}
\caption{Two-dimensional pion and kaon PID distributions in the original $n\sigma_{\pi}-m^{2}$ space and in the transformed $x-y$ space. Panels (a) and (c) show the smeared distributions before the transformation in the transverse momentum intervals $2.0 < p_T < 2.2$ and $2.8 < p_T < 3.0$~GeV/$c$, respectively. Panels (b) and (d) show the corresponding distributions after the scale, shift, and rotation transformation for the same $p_T$ intervals. The increasing overlap in the original PID space at high $p_{T}$ motivates the use of the combined two-dimensional information.}
\label{fig:fig2}
\end{figure}

\begin{figure}[htbp]
\begin{center}
\includegraphics[width=0.6\linewidth]{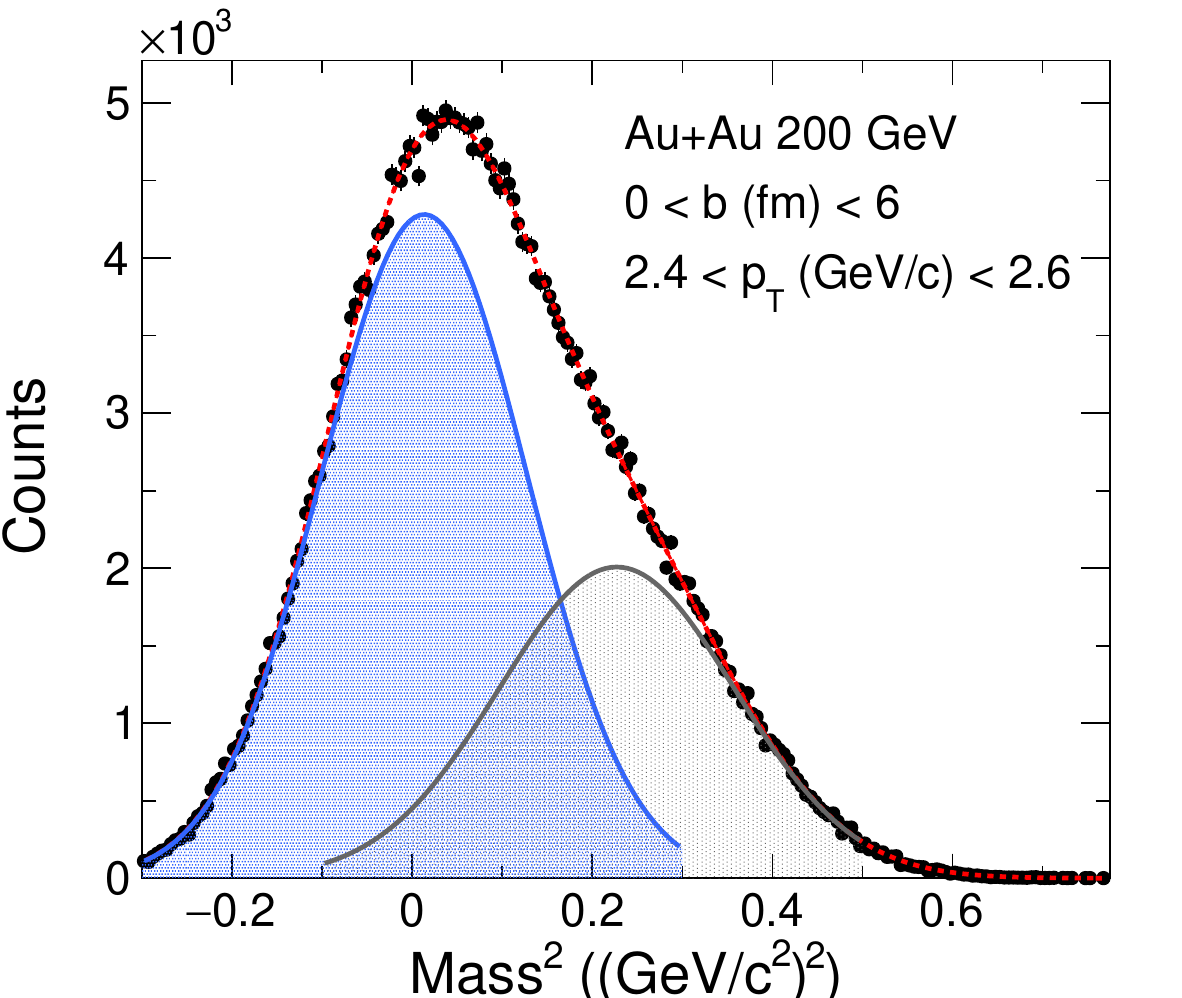}
\caption{Example of the reference one-dimensional PID procedure for charged pions and kaons in the transverse momentum interval $2.4 < p_T < 2.6$~GeV/$c$. Tracks are first selected with $|n\sigma_{\pi}|<$3 , after which the resulting $m^{2}$ distribution is fitted with pion and kaon Gaussian components. The red dashed line represents the total fit, while the individual curves show the pion and kaon components extracted from the fit, respectively.}
\label{fig:fig3}
\end{center}
\end{figure}

The left panels of Fig.~\ref{fig:fig2} show the two-dimensional distributions of $m^{2}$ versus $n\sigma_{\pi}$ in the transverse momentum intervals 2.0$<p_T<$2.2 GeV/$c$ (top) and 2.8$<p_T<$3.0 GeV/$c$ (bottom). These distributions clearly illustrate the increasing overlap between pion and kaon bands with increasing $p_T$.
For a quantitative reference, we define a one-dimensional PID procedure in which tracks are first selected with $|n\sigma_{\pi}|<$3. Within each $p_{T}$ interval, a one-dimensional $m^{2}$ distribution is then constructed from the selected tracks, as shown by the black solid circles in Fig.~\ref{fig:fig3}, and fitted with the sum of pion and kaon Gaussian components. Particle yields are subsequently extracted by integrating the fitted function within a window defined by the mean $\pm3\sigma$ of each peak. 
Fig.~\ref{fig:fig3} presents a representative example of this reference approach for pion and kaon identification in the $p_T$ range of 2.4-2.6 GeV/$c$. While this method performs adequately at low transverse momentum, it begins to exhibit pronounced limitations in the higher $p_T$ region. In particular, the kaon component increasingly overlaps with the dominant pion distribution and becomes progressively more difficult to separate, making reliable particle identification increasingly difficult within this one-dimensional framework.

To improve the separation of the pion and kaon components, the original $n\sigma_{\pi} - m^{2}$ coordinates are transformed through scaling, translation, and rotation, as illustrated in the right panels of Fig.~\ref{fig:fig2}. 
The transformation defines a new coordinate system, $x(n\sigma_{\pi},m^{2})$ and $y(n\sigma_{\pi},m^{2})$, in which the effective widths of the particle distributions along two axes become comparable and the dominant pion-kaon separation is oriented approximately along the $x$ direction. This representation facilitates a more stable decomposition of the two species in regions of strong PID overlap.


The transformation consists of three sequential steps $\textendash$ scaling, shifting, and rotation $\textendash$ and is defined by Eqs.~\ref{equ:fscale},\ref{equ:shift_x},\ref{equ:shift_y},\ref{equ:angle_rotate},\ref{equ:rotation}.
First, a scale factor $f_{\rm scale}$ is introduced to normalize the widths of the distributions along the $n\sigma_{\pi}$ and $m^{2}$ axes. The scaling step equalizes the effective resolution of the two variables, ensuring that neither dimension dominates the particle separation in the transformed space. Next, a linear translation is applied to shift the centroid of the pion distribution to the origin of the coordinate system. The shift parameter, $\mu_{\pi}(n\sigma_{\pi})$ and $\mu_{\pi}(m^{2})$, correspond to the mean values of the pion $n\sigma_{\pi}$ and $m^{2}$ distributions, respectively. These parameters are obtained from the parameterized $p_T$-dependent functions shown in Fig.~\ref{fig:fig1}.
Following the scaling and shifting steps, a rotation is applied to further optimize the alignment of the particle bands. The rotation angle $\alpha$ is determined from the relative orientation of the kaon band with respect to the pion band in the two-dimensional $m^{2}-n\sigma_{\pi}$ space. A counterclockwise rotation by angle $\alpha$ is then performed, such that the kaon band becomes aligned parallel to the pion band along the $x$ axis.

As a result of these transformations, the pion and kaon distributions become approximately aligned in the transformed coordinate system, with their centroid separation oriented predominantly along the $x$ direction.. 
Since the transformation is linear and applied uniformly on a track-by-track basis, it preserves event-by-event correlations and does not introduce artificial azimuthal modulations, making it suitable for subsequent yield and flow analyses.

\begin{figure}[htbp]
\centering
\centerline{\includegraphics[width=0.6\linewidth]{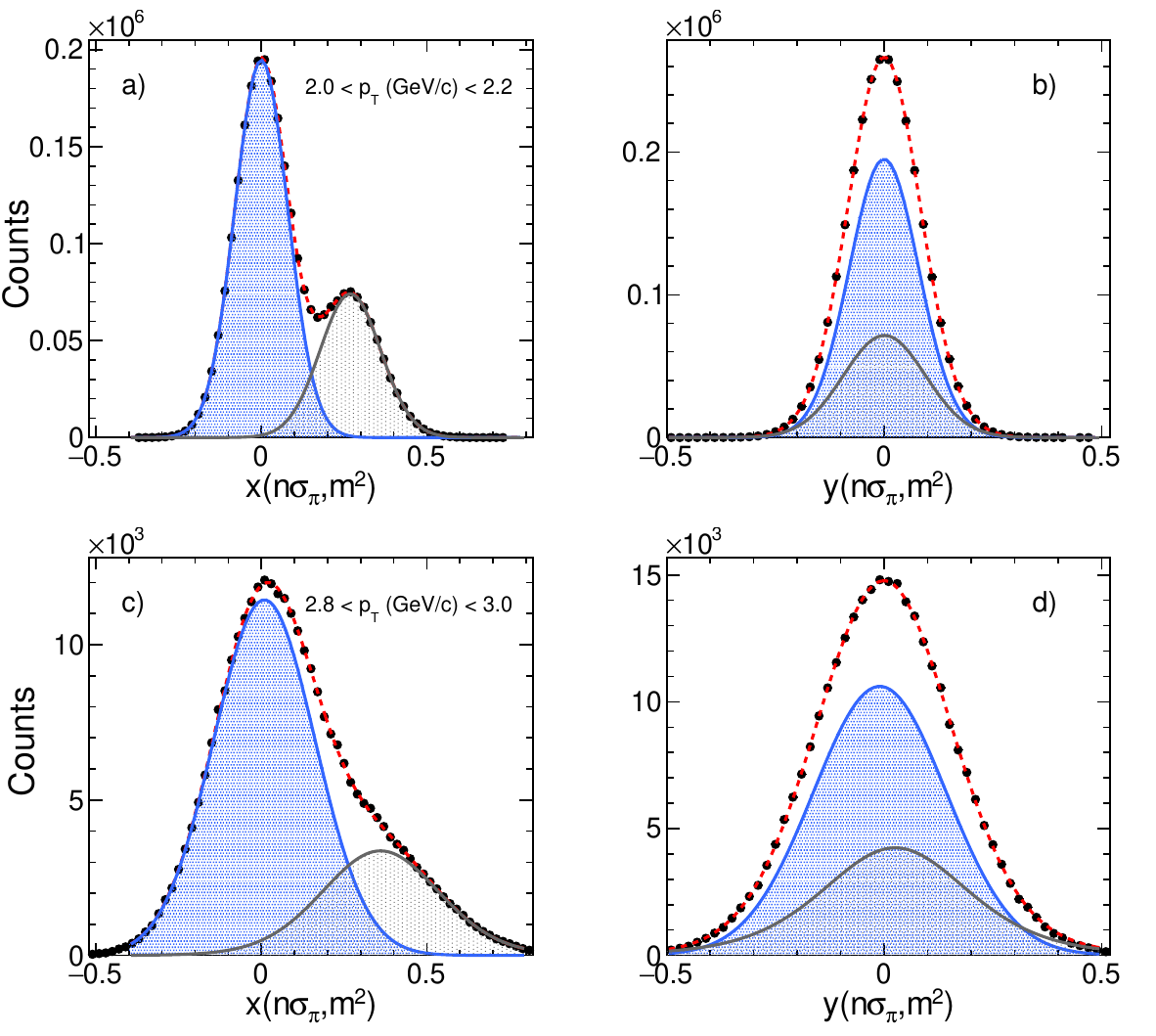}}
\caption{Projections of the transformed two-dimensional pion and kaon distributions onto the $x$ and $y$ axes for $2.0 < p_T < 2.2$~GeV/$c$ [(a), (b)] and $2.8 < p_T < 3.0$~GeV/$c$ [(c), (d)]. The dashed curves denote the nominal Student-$t$ fit, while the individual curves represent the pion and kaon components.}
\label{fig:fig4}
\end{figure}

\begin{equation}
    f_{\rm scale}=\frac{\sigma_{\pi}(n\sigma_{\pi})}{\sigma_{\pi}(m^{2})}
    \label{equ:fscale}
\end{equation}
\begin{equation}
    x'=\frac{n\sigma_{\pi}-\mu_{\pi}(n\sigma_{\pi})}{f_{\rm scale}}
    \label{equ:shift_x}
\end{equation}
\begin{equation}
    y'=m^{2}-\mu_{\pi}(m^{2})
    \label{equ:shift_y}
\end{equation}
\begin{equation}
    \alpha=-{\rm tan}^{-1}\left[\frac{\mu_{K}(m^{2})-\mu_{\pi}(m^{2})}{(\mu_{K}(n\sigma_{\pi})-\mu_{\pi}(n\sigma_{\pi}))/f_{\rm scale}}\right]
    \label{equ:angle_rotate}
\end{equation}
\begin{equation}
    \begin{pmatrix}
x(n\sigma_{\pi},m^{2})\\
y(n\sigma_{\pi},m^{2})
\end{pmatrix}
=
\begin{pmatrix}
{\rm cos}(\alpha)&- {\rm sin}(\alpha)\\
{\rm sin}(\alpha)&{\rm cos}(\alpha)
\end{pmatrix}
\begin{pmatrix}
x'\\
y'\\
\end{pmatrix}
\label{equ:rotation}
\end{equation}

As shown in the right panels of Fig.~\ref{fig:fig4}, the transformation leads to two notable improvements in the particle distributions. First, for each particle species, the widths of the distributions along the $x$ and $y$ axes become approximately identical. Second, the pion and kaon distributions are well aligned along the horizontal axis in the transformed coordinate system.
These features allow the transformed distributions to be described with a relatively simple factorized parameterization and improve the numerical stability of the pion–kaon decomposition in regions where the original PID responses strongly overlap.


In this analysis, the transformed pion and kaon distributions are described using a factorized two-dimensional Student-$t$ parameterization.  The choice of the Student-$t$ function follows the fitting strategy used in experimental applications, where additional flexibility may be useful. In the present Monte Carlo study, the detector response is Gaussian by construction. Therefore, the use of the Student-$t$ function is not interpreted as evidence for non-Gaussian detector tails in the simulated sample. To test the dependence on the fitting form, the complete extraction is repeated using a factorized two-dimensional Gaussian parameterization. 
Because the detector response in the present benchmark is Gaussian by construction, the Student-$t$ versus Gaussian comparison tests the sensitivity to the fitting function rather than the effect of genuine non-Gaussian detector response tails. A quantitative treatment of such tails would require a detector-specific response model and is beyond the scope of the present controlled closure study.

The one-dimensional Student-$t$ function is characterized by three parameters: the mean($\mu$), the width($\sigma$), and the degrees-of-freedom parameter($\nu$). When extended to two dimensions, an additional normalization parameter is introduced for each particle species, as summarized in Eq.~\ref{equ:2d_student_t}.

In total, the combined two-dimensional fit function used to describe the pion and kaon distributions is given in Eq.~\ref{equ:2d_totalfit}. Because of the relatively large parameter space and correlations among the fit parameters, the fitting procedure is performed iteratively. Approximate peak positions and widths obtained from preliminary fits are used only as initial values for the minimization and are not fixed during the subsequent fitting procedure. To improve the stability of the fit in regions where the pion and kaon distributions strongly overlap, a subset of the shape parameters is constrained within physically reasonable ranges, while the particle yields remain free parameters. The fitted parameters obtained from each iteration are then used as the starting values for the subsequent iteration, and the procedure is repeated until stable convergence is achieved.

\begin{equation}
    f_{x(y)}=
\frac{\Gamma(\frac{\nu_{x(y)}+1}{2})}
{\Gamma(\frac{\nu_{x(y)}}{2})\sqrt{\pi\nu_{x(y)}\sigma^{2}_{x(y)}}}
\begin{bmatrix}
1+\frac{1}{\nu_{x(y)}}
(\frac{x(y)-\mu_{x(y)}}{\sigma_{x(y)}})^{2}
\end{bmatrix}
^{-\frac{\nu_{x(y)}+1}{2}}
\label{equ:student_t}
\end{equation}

\begin{equation}
    f_{particle} = N_{0}\times f_{x,particle} \times f_{y,particle}
    \label{equ:2d_student_t}
\end{equation}

\begin{equation}
    f_{fit}=f_{\pi}+f_{K}
    \label{equ:2d_totalfit}
\end{equation}

After applying the transformation described above, the yields of charged pions and kaons are extracted by projecting the two-dimensional distribution onto the $x$ axis. As shown in the left panels of Fig.~\ref{fig:fig4}, the resulting one-dimensional distributions are well described by the multi-component Student-$t$ function used in the fit. The pion and kaon peaks are clearly separated, enabling a stable and reliable extraction of particle yields over the full transverse momentum range considered.
For each $p_T$ interval, the particle yield is evaluated by integrating the corresponding fitted component within a $\pm3\sigma$ window centered around the corresponding particle peak.

The right panels of Fig.~\ref{fig:fig4} show the projection of the same two-dimensional distribution onto the $y$ axis. The resulting distributions are narrow and approximately symmetric, indicating that, after the shift and rotation, both pion and kaon distributions are effectively aligned along the horizontal direction in the transformed coordinate system. 
The $y$-axis projections provide an additional visualization of the transformed distributions and their approximate alignment in the transformed coordinate system. This alignment facilitates a stable numerical decomposition of the pion and kaon components in regions of strong PID overlap.

To verify that the extracted performance does not depend on the particular fitting function, the complete two-dimensional analysis is repeated using a two-dimensional Gaussian parameterization. The Gaussian and Student-$t$ results are found to be consistent over the studied $p_{T}$ range.


\section{Results and discussion}
\label{sec:res}


\begin{figure}[htbp]
\begin{center}
\includegraphics[width=0.6\linewidth]{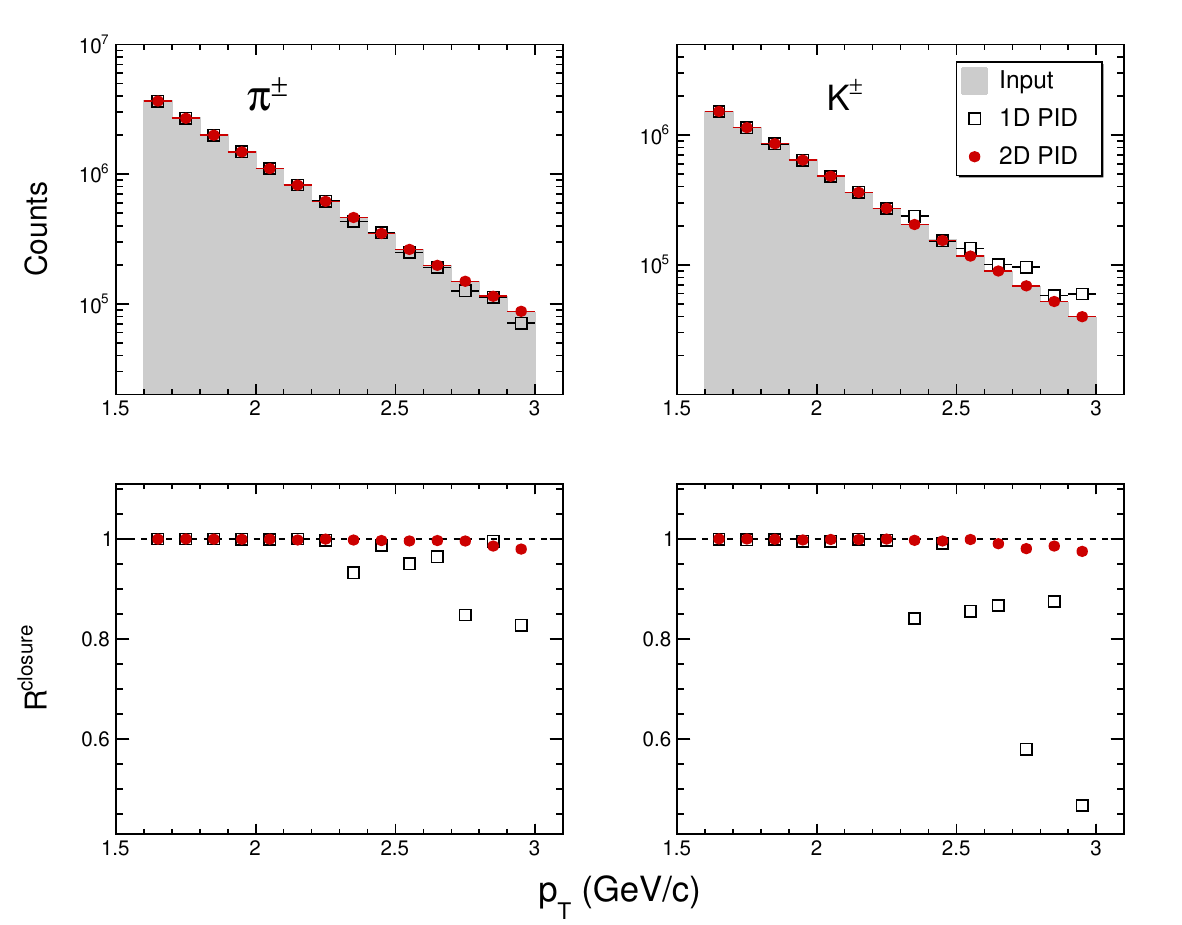}
\caption{
Transverse momentum ($p_{T}$) distribution of reconstructed charged pions (left) and charged kaons (right) yields. The AMPT input is compared with the reference one-dimensional $m^{2}$ fit procedure and the two-dimensional Student-$t$ procedure. The lower panels show the corresponding yield closure ratio, $R^{\textrm{closure}} = N^{\textrm{extracted}} / N^{\textrm{input}}$. The same AMPT input yield, evaluated after the common kinematic selections and before any PID selection, is used as the denominator for all reconstruction methods.
}
\label{fig:fig5}
\end{center}
\end{figure}

Figure~\ref{fig:fig5} presents a direct quantitative comparison between the reference one-dimensional PID procedure and the two-dimensional shift and rotation procedure. The AMPT input yields are used as the common truth reference. To quantify the reconstruction performance, we define the yield closure ratio:
\begin{equation}
    R^{\text{closure}}(p_T) = \frac{N^{\text{extracted}}(p_T)}{N^{\text{input}}(p_T)},
    \label{equ:ratio}
\end{equation}
Because the denominator is defined before any PID selection, $R^{\rm closure}$ for the reference one-dimensional procedure includes the efficiency loss introduced by the $|n\sigma_{\pi}|<$ 3 preselection. This definition is intentional, because the comparison is designed to evaluate the performance of the complete PID procedures rather than only the subsequent fit.
The red solid circles denote the yields obtained with the two-dimensional Student-$t$ procedure, while the black open squares represent the  results from the reference one-dimensional procedure and the gray band indicates the AMPT input. At relatively low $p_{T}$, where the pion and kaon PID responses are well separated, both procedures reproduce the input yields within the corresponding uncertainties. As $p_{T}$ increases, the overlap between the pion and kaon PID responses becomes more pronounced, and the reference one-dimensional procedure exhibits a progressively larger deviation from the input, particularly for kaons. In contrast, the two-dimensional procedure maintains good yield closure over the full $p_{T}$ range studied, with $R^{\rm closure}$ remaining within approximately 2\% of unity up to $p_{T} = $ 3 GeV/$c$.
As a fit-model cross-check, the extraction is repeated using a factorized two-dimensional Gaussian parameterization. The resulting yields are consistent with those obtained using the nominal Student-$t$ parameterization within the corresponding uncertainties. This agreement indicates that the improved reconstruction performance is not driven by the specific choice of fitting function, but rather by the simultaneous use of the $n\sigma_{\pi}$ and $m^{2}$ PID information in the two-dimensional procedure. This enhanced high-$p_{T}$ capability is essential for precision measurements that rely on identified hadrons, particularly for observables with strong transverse momentum dependence, such as anisotropic flow and number-of-constituent-quark scaling studies.


\begin{figure}[htbp]
\centering
\centerline{\includegraphics[width=0.6\linewidth]{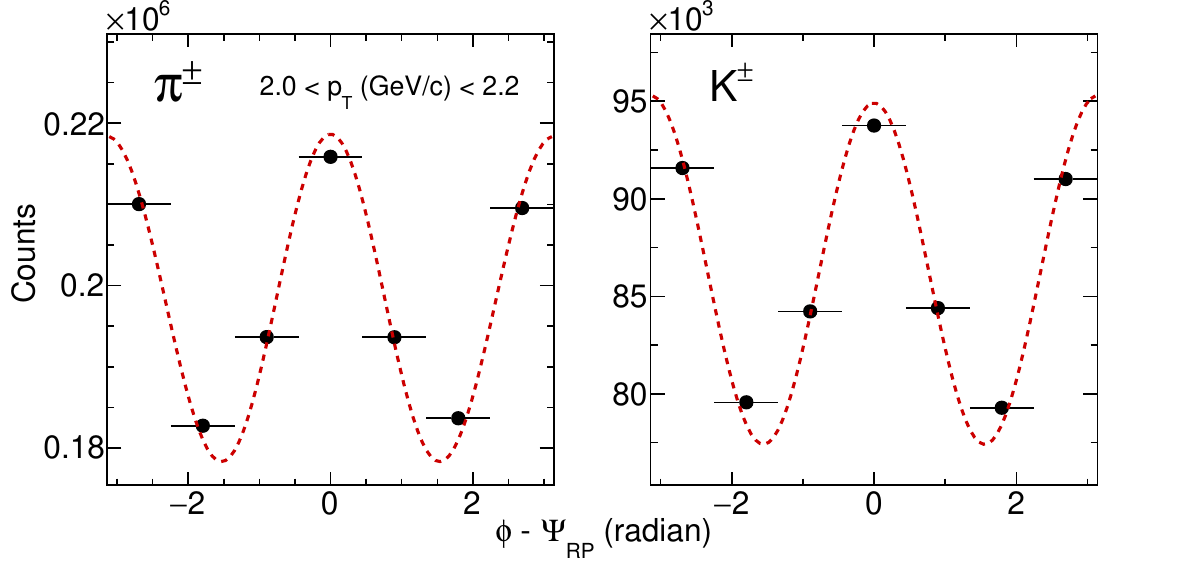}}
\caption{Azimuthal yield distributions relative to the reaction plane, $\phi-\Psi_{\rm RP}$, for charged pions (left) and kaons (right) in $2.0<p_T<2.2$~GeV/$c$. The symbols show the yields reconstructed with the two-dimensional PID procedure, while the dashed curves show the Fourier fits used to extract $v_2$.}
\label{fig:fig6}
\end{figure}

\begin{figure}[htbp]
\begin{center}
\includegraphics[width=0.6\linewidth]{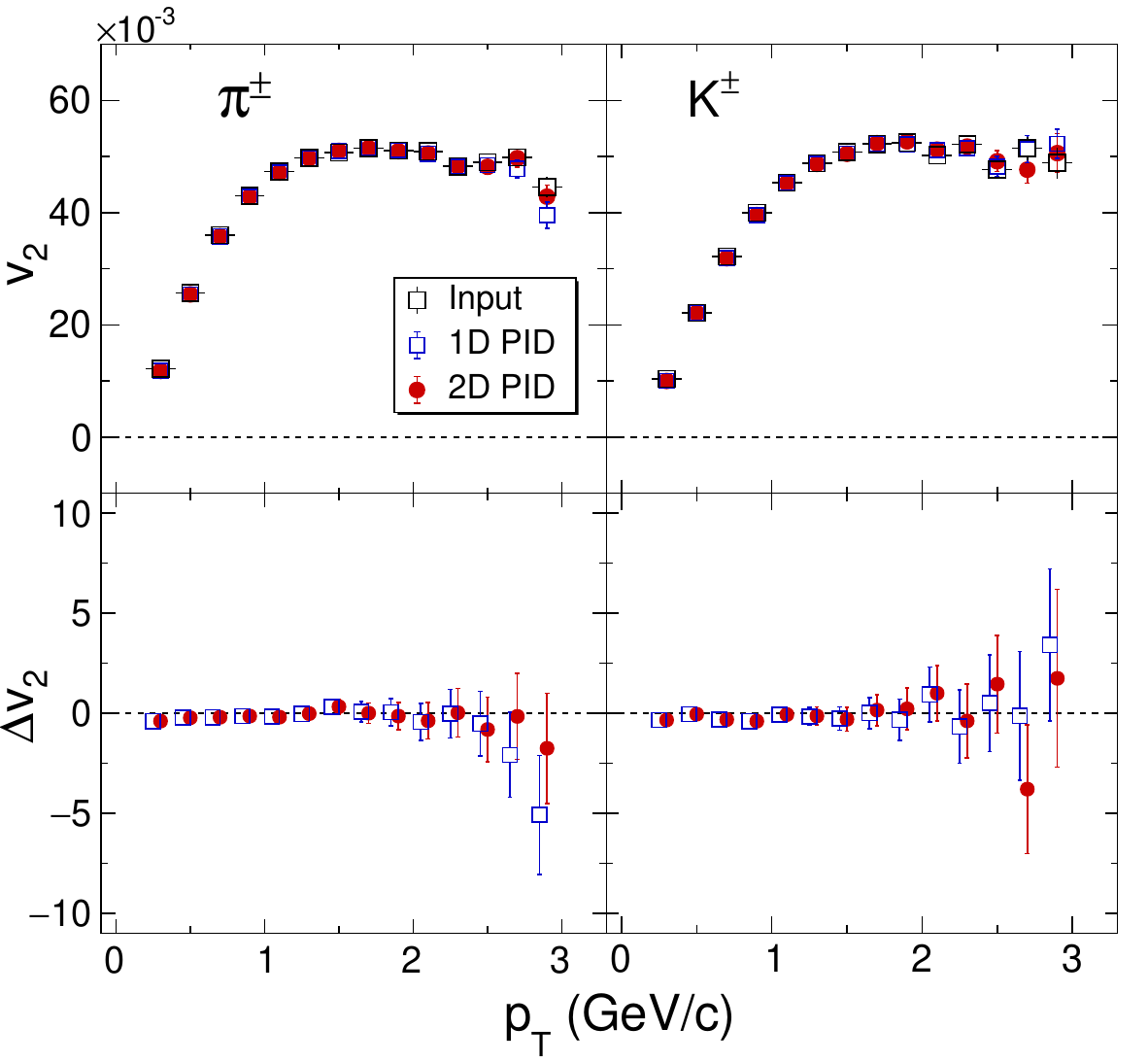}
\caption{Transverse momentum ($p_T$) dependence of the reaction-plane elliptic flow coefficient $v_2$ for charged pions (left) and kaons (right). The upper panels compare the AMPT input with the results reconstructed using the reference one-dimensional PID procedure and the two-dimensional shift and rotation procedure. The lower panels show the difference between the extracted and input $v_{2}$ values, providing a direct measure of the PID induced deviation from the AMPT input. Both reconstruction procedures are consistent with the input $v_{2}$ within the statistical uncertainties over the studied $p_{T}$ range.}
\label{fig:fig7}
\end{center}
\end{figure}

To investigate whether the PID reconstruction introduces a bias in anisotropic flow measurements, we perform a  reaction-plane $v_{2}$ closure test~\cite{Voloshin:2008dg,Poskanzer:1998yz,STAR:2025owm,STAR:2026axo,Zhou:2024cte}, and the extracted results are directly compared with the known input values from the AMPT model.
In the AMPT simulation, the true event geometry is explicitly available, allowing the input elliptic flow to be calculated with respect to the reaction plane according to $v_{2} = \langle \rm cos[2(\phi - \Psi_{\rm RP})] \rangle$, where $\phi$ is the azimuthal angle of the particle and $\Psi_{\rm RP}$ denotes the reaction plane angle. The average is taken over all particles in all events. In the present setup, the reaction plane is fixed at $\Psi_{\rm RP}=0$, which simplifies the evaluation of the input $v_2$.

For each $p_T$ interval, the relative azimuthal angle $\phi - \Psi_{\rm RP}$ is divided into seven uniform bins. The pion and kaon yields are reconstructed independently in each azimuthal bin using both the reference one-dimensional and the two-dimensional PID procedures. The resulting azimuthal yield distributions are fitted with the following Fourier expansion:
\begin{equation}
    \frac{dN}{d(\phi-\Psi_{\rm RP})} = N_{0} [1 + 2  v_{1} {\rm cos}(\phi-\Psi_{\rm RP}) + 2 v_{2} {\rm cos}(2(\phi-\Psi_{\rm RP}))].
\end{equation}
where $v_1$ and $v_2$ represent the first two harmonic flow coefficients. Contributions from higher-order harmonics are expected to be negligible within the statistical precision of the present study and are therefore not included in the fit.
Figure~\ref{fig:fig6} shows representative azimuthal yield distributions and the corresponding fits for charged pions and kaons in the transverse momentum interval $2.0<p_T<2.2$~GeV/$c$.

Figure~\ref{fig:fig7} compares the reconstructed reaction plane $v_{2}$ values obtained with the reference one-dimensional and two-dimensional PID procedures with the known AMPT input. For both charged pions and kaons, the results obtained with the two PID procedures are consistent with the input values within the statistical uncertainties over the full $p_{T}$ range studied. The corresponding differences, $\Delta v_{2} = v_{2}^{\rm extracted} - v_{2}^{\rm input}$, are shown in the lower panels of Fig.~\ref{fig:fig7}. 
No statistically significant deviation from zero is observed for either procedure. Although the statistical uncertainties increase toward the highest $p_{T}$ values, both the reference one-dimensional and two-dimensional results remain consistent with the AMPT input within uncertainties, with no indication of a systematic PID-induced bias in the extracted $v_{2}$. 
Repeating the $v_{2}$ extraction with the alternative $\mu \pm 2\sigma$ integration window yields results consistent with the nominal $\mu \pm 3\sigma$ extraction within the statistical uncertainties. The two-dimensional Gaussian parameterization likewise produces no statistically significant change in the extracted $v_{2}$.

It is noteworthy that the degradation of the yield closure observed for the reference one-dimensional PID procedure does not lead to a comparably large bias in the extracted $v_{2}$. This behavior is consistent with a PID efficiency loss that is approximately independent of the azimuthal angle relative to the reaction plane: in that case, the overall yield normalization is modified, while the harmonic modulation from which $v_{2}$ is extracted is largely preserved. Within the statistical precision of the present study, both PID procedures therefore provide a consistent reconstruction of the input reaction plane $v_{2}$, although the two-dimensional procedure provides substantially improved yield closure at high $p_{T}$.

\section{Summary}
\label{sec:sum}

In summary, we have quantitatively benchmarked a two-dimensional pion-kaon particle identification procedure based on a scale, shift, and rotation of the combined $n\sigma_{\pi}$ and $m^{2}$ information. The detector response is constructed using $p_{T}$-dependent PID parameters extracted from STAR data and applied to AMPT particles through Gaussian smearing.
The AMPT true particle identities are not used during the PID reconstruction and are retained only for the final closure comparison, allowing the reconstruction performance to be quantified directly.
The two-dimensional procedure is compared directly with a reference one-dimensional method consisting of an $|n\sigma_{\pi}|<$ 3 preselection followed by a Gaussian fit to the $m^{2}$ distribution. Both procedures reproduce the input yields well when the pion and kaon PID responses are clearly separated. As the PID overlap increases toward higher $p_{T}$, however, the reference one-dimensional procedure develops large deviations from the known AMPT yields, particularly for kaons, whereas the two-dimensional procedure maintains the yield closure ratio $R^{\rm closure}$ within approximately 2\% of unity up to $p_{T} = $ 3 GeV/$c$.

The dependence of the two-dimensional yield extraction on the fitting form is examined by replacing the nominal Student-$t$ parameterization with a factorized two-dimensional Gaussian fit. The resulting yields are consistent within the corresponding uncertainties, indicating that the improved yield closure is associated primarily with the joint two-dimensional treatment of the PID information rather than with the specific fitting function.

The influence of the PID reconstruction on elliptic flow has also been investigated using a reaction-plane $v_{2}$ closure test. Within the statistical precision of the present study, both the reference one-dimensional and two-dimensional procedures reproduce the AMPT input $v_{2}$ over the full $p_{T}$ range considered. This finding shows that the reduced yield closure of the one-dimensional procedure does not necessarily translate into a comparable bias in $v_{2}$, consistent with a PID efficiency loss that is approximately independent of the azimuthal angle relative to the reaction plane.

The general strategy may also be adapted to future experiments such as the Cooling Storage Ring External-target Experiment (CEE) at HIRFL/HIAF, the Nuclotron-based Ion Collider Facility (NICA) at JINR, and the Compressed Baryonic Matter (CBM) experiment at FAIR~\cite{Lu:2016htm, Liu:2023xhc, Zhang:2023hht, Wu:2025dev, Kisiel:2020spj, MPD:2025jzd, CBM:2025voh}. 
These facilities will explore different regions of the QCD phase diagram with varying collision energies and system sizes, where precise particle identification across extended transverse momentum ranges will be crucial. 
However, its quantitative momentum reach must be reassessed using the detector-specific PID response and resolution of each experiment.
The two-dimensional shift and rotation approach provides a flexible framework that can be tuned to the specific detector characteristics and PID requirements of these experiments, thereby enhancing their capability to extract clean signals of identified hadrons for studying the properties of dense nuclear matter.

\acknowledgments
We thank Prof. Shusu Shi for useful discussions. This work was supported by the Doctoral Scientific Research Foundation of Pingdingshan University (PXY-BSQD-2023016), the Natural Science Foundation of Henan under contract No. 252300420921, the National Key Research and Development Program of China under Contract No. 2024YFA1610700.

\bibliography{ref.bib}


\end{document}